\documentclass[prl,twocolumn,showpacs,preprintnumbers,amsmath,amssymb]{revtex4}

\usepackage{graphicx}
\usepackage{dcolumn}
\usepackage{bm}

\begin{document}

\title{Three-dimensional fast electron transport for ignition-scale
inertial fusion capsules}

\author{J.J. Honrubia}
\email{javier.honrubia@upm.es}
\affiliation{ETSI Industriales, Universidad Polit$\acute{e}$cnica, Madrid, Spain}

\author{J. Meyer-ter-Vehn}
\email{juergen.meyer-ter-vehn@mpq.mpg.de}
\affiliation{Max-Planck-Institut f\"ur Quantenoptik, Garching, Germany}

\date{\today}

\begin{abstract}

Three-dimensional (3D) hybrid PIC simulations are presented to
study electron energy transport and deposition in a full-scale
fast ignition configuration. Multi-prong core heating close to
ignition is found when a few GA, few PW beam is injected.
Resistive beam filamentation in the corona seeds the 3D current
pattern that penetrates the core. Ohmic heating is important
in the low-density corona, while classical Coulomb deposition heats
the core. Here highest energy densities (few Tbar at 10 keV)
are observed at densities above 200 g/cm$^3$. Energy coupling
to the core ranges from 20 to 30\%; it is enhanced by beam collimation
and decreases when raising the beam particle energy from 1.5 to 5.5 MeV.

\end{abstract}


\maketitle

Fast ignition of inertial fusion targets \cite{tabak1, tabak2}
offers a promising
alternative to the standard scheme of central hot spot ignition
\cite{lindl,MtV2}. Separating fuel compression from hot spot heating
is expected to reduce compression and symmetry requirements significantly.
Here we study Tabak's original proposal \cite{tabak1} to use
a laser-driven electron beam to ignite the compressed core.
Atzeni \cite{atzeni,MtV2} estimates that an energy of about 20 kJ
is required to ignite 300 g/cm$^3$ Deuterium-Tritium (DT) fuel when
deposited in 20 ps on a 20 $\mu$m spot radius with 0.6 g/cm$^2$
stopping range. This implies a giga-ampere (GA), peta-watt (PW) pulse
of 1 MeV electrons.
Little cones may be used to generate such beams and to guide
them through the plasma corona \cite{kodama,LLE}.
Recent experiments on cone guiding with 300 J, 0.5 PW laser
pulses have demonstrated excellent (20 -30\%) energy coupling
from laser to core \cite{kodama, kodama2002}.
For full-scale fast ignition, a 100 kJ, multi-PW beam is
required including energy coupling.
It will carry a few GA current, 10$^4$ times larger than the
Alfv$\acute{e}$n current that may limit transport
due to magnetic self-interaction. Here the beam has to be
transported over a distance of 100 - 200 $\mu$m between
cone tip and core through a high-gradient plasma profile.
This is the topic of the present paper.

In plasma, the beam current is compensated by return currents,
which suppress the magnetic fields. But this beam is subject to
filamentation instability. For collisionless plasma, linear growth
rates have been studied in \cite{silva,bret}, and particle-in-cell
(PIC) simulation was used to trace the nonlinear evolution
\cite{Honda2000,Sentoku2003}. This applies to low plasma densities,
comparable to the beam density ($n_{beam}/n_{plasma} \ge 0.1$)
but not to the high plasma densities considered in this paper.
Full-scale PIC simulations are not yet feasible. They
should include collisions and plasma resistivity to properly
describe the return currents. Here we use a hybrid model
adequate for describing self-magnetized transport in high-density
fuel. It treats only the relativistic beam electrons by PIC and
models the background plasma by the return current density {\bf j$_r$},
tied to the electric field {\bf E}=$\eta${\bf j$_r$} by Ohm's law
with resistivity $\eta$. Maxwell's equations are used in the form
$\nabla\times${\bf B}=$\mu_0${\bf j} and
$\nabla\times${\bf E}=$-\partial${\bf B}/$\partial$t, where
{\bf j}={\bf j$_b$}+{\bf j$_r$} is the sum of beam and return current
density. The displacement current and charge separation effects can
be neglected since in this high-density environment relaxation
times and Debye lengths are much smaller than the sub-picosecond
and micrometer scales of the resistive filamentation
investigated here. The beam deposits energy into plasma electrons
in two ways: by direct classical Coulomb deposition and via
return current ohmic heating with power density $\eta j_r^2$.
Electrons and ions are coupled by thermal energy transfer.
A plasma density profile constant in time with equal
electron and ion number densities is assumed.
This model was proposed by Bell \cite{bell} and
further developed by Davies \cite{davies} and Gremillet et al.
\cite{gremillet}. First three-dimensional (3D) simulations based
on this model were published in \cite{gremillet}, showing 3D
resistive filamentation. Gremillet et al. also derived the
linear growth rates, now depending on resistivity $\eta$. The
present version of the model is described in more detail in
\cite{honrubia1}.

Recently, two-dimensional simulations of the cone-guided target
experiment \cite{kodama, kodama2002} were published by Campbell et al.
\cite{campbell} and Mason \cite{mason}. Based on different
hybrid codes, they could reproduce the measured core heating
of 800 eV. Using the present model, we have obtained similar
2D results \cite{honrubia2}; comparisons with the linear theory
of the resistive filamentation instability are found in \cite{MtV}.
Here we present first three-dimensional simulations of electron
transport and deposition in the high-density part of a fast
ignition target. It is shown how 3D beam filamentation seeded
in the corona leads to multi-hot-spot heating of the core close
to DT ignition. The present study is motivated by the next generation
of high-power facilities to demonstrate fast ignition
\cite{dunne,firex}.

\begin{figure}
\begin{center}
\includegraphics[width=.47\textwidth]{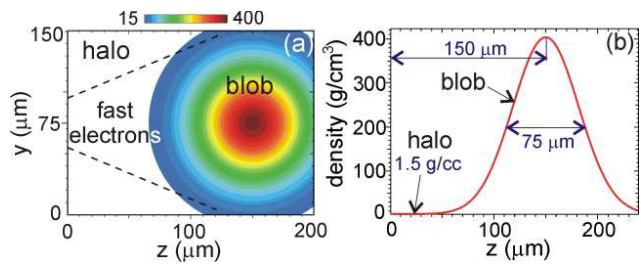}
\caption{\label{fig:1} Central cut through imploded
target configuration: (a)  isocontours of density in g/cm$^3$,
(b) density profile at $y$=75 $\mu m$.}
\end{center}
\end{figure}

{\it Simulation parameters.}
The simulated configuration is shown in figure~\ref{fig:1}.
It consists of 0.2 mg DT fuel compressed into a spherical
blob of 400 g/cm$^3$ peak density and 75 $\mu$m diameter
(full width at half maximum, FWHM); it sits on a
density pedestal of 1.5 g/cm$^3$ (the {\it halo}).
A beam of fast electrons is injected from the left
at $z=0$. We imagine that it emerges from the tip of
a cone at this position. The cone itself and the
laser pulse generating the beam inside the cone are
not simulated here. We rather model the injected
beam in form of directed Gaussian electron
distributions in radius and time with a spot
radius of 20 $\mu$m and a duration of 10 ps,
both at FWHM. The pulse has a power of 6 PW,
a total energy of 60 kJ ($\approx$30 kJ within FWHM)
and is centered at 7 ps.
The energy distribution of the beam electrons is
assumed to be 1-D relativistic Maxwellian with
temperatures depending on the local laser
intensity $I$, also assumed Gaussian in radius
and time, by the ponderomotive scaling formula
$T_b\approx f\cdot m_ec^2\,[(1+I\lambda^2/13.7\,$GW$)^{1/2}-1]$,
where $\lambda$ is the wavelength.
PIC simulations \cite{pukhov} give front factors
$f \approx 1 - 3$, depending on the scale-length
of the plasma in which the electrons are accelerated.
For cone-guided fast ignition with 10 ps pulse
durations and electron acceleration along the
cone surface \cite{Sentoku2004}, the factor is
expected to be larger than $f\approx 1$ which
applies to sharp surfaces. Here we consider
different cases with mean electron kinetic energies
(averaged over the FWHM of the distributions in
radius and time) in the range of $\langle E\rangle =$
1.5 - 5.5 MeV; 2.5 MeV is taken as a reference value.
This mean energy corresponds to a laser irradiance
of $1.5\times10^{20}\,W/cm^2$ (FWHM) at 0.35 $\mu$m,
assuming laser-to-electron transfer of 50\%
and beam compression by a factor of 3 due to
geometrical cone convergence, in agreement with
the results reported in \cite{kodama,kodama2002,
Sentoku2004,Kodama2004a}. The initial angular
distribution of fast electrons is obtained as in
\cite{honrubia3}.
Electrons with energy $E=(\gamma-1)m_ec^2$ are
injected with a randomly chosen half-angle between
$0$ and $\tan^{-1}[h\sqrt{2/(\gamma-1)}]$.
The parameter $h$ is used to adjust the initial
beam opening half-angle as 22.5$^\circ$ (FWHM),
consistent with the cone experiment \cite{kodama,
kodama2002} and the simulations in \cite{campbell,
mason}.

The imploded fuel configuration shown in figure~\ref{fig:1}
has been scaled from that reported in \cite{campbell,mason}.
The main parameters of this configuration, the peak density
of 400 g/cm$^3$, the distance of 150~$\mu$m from cone
tip, and the initial plasma temperature, have been adapted
from implosion simulations of cone targets with the code
SARA-2D \cite{honrubia4}. The plasma resistivity
depends on the temperature distribution. The SARA-2D
simulations indicate temperatures in the range of
300 eV to 1 keV. For simplicity, a uniform initial
DT temperature of 500 eV is taken here, which sets
the initial resistivity to a level of $10^{-8}$ $\Omega$m.
Concerning the numerical parameters, we have chosen a
cell width of 1 $\mu$m in each coordinate, a time step
of 3 fs, and a total number of $3.6\times10^{7}$
particles injected over the time interval of 0 - 14 ps.
Free boundaries have been used in all simulations. Classical
Spitzer resistivity is chosen for the DT plasma, and
MPQeos tables \cite{kemp} are used to compute electron
and ion temperatures from the deposited energy.

\begin{figure}
\begin{center}
\includegraphics[width=.45\textwidth]{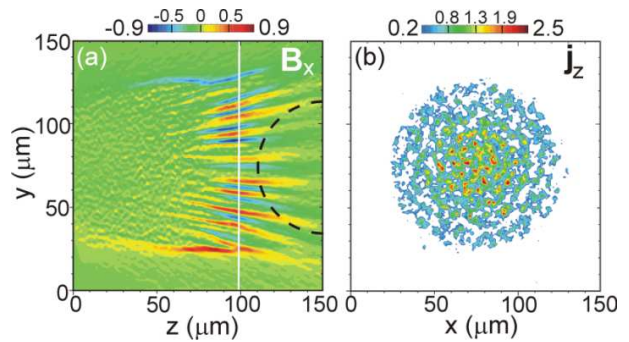}
\caption{\label{fig:2}
(a) Central longitudinal cut of magnetic field $B_x$ in kT
generated by electrons with a mean kinetic energy of 2.5 MeV
at the peak of the pulse,
(b) perpendicular cut of beam current density $j_z$
in units of 10$^{14}$ A/cm$^2$ at $z$=98 $\mu$m. Plasma
densities higher than 200 g/cm$^3$ are located inside
the dashed circle.}
\end{center}
\end{figure}

\begin{figure}
\begin{center}
\includegraphics[width=.45\textwidth]{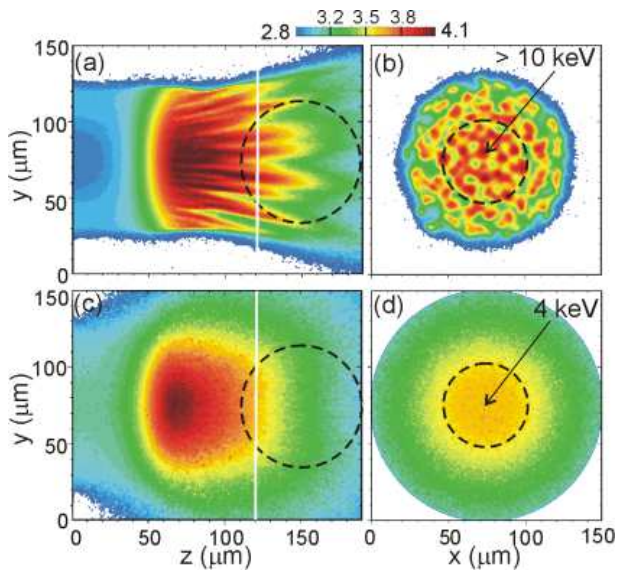}
\caption{\label{fig:3} Ion temperature
of DT in units of $\log_{10}(T_i/$eV)
at the end of the pulse of electrons with
a mean kinetic energy of 2.5 MeV.
(a) longitudinal cut at $x=75$ $\mu$m, (b) transverse
cut at $z=120 $ $\mu$m. Beam-generated fields
have been artificially suppressed in (c)
and (d). Plasma densities higher than 200 g/cm$^3$
are located inside the dashed circle.}
\end{center}
\end{figure}

{\it Results.}
The injected current of 3.5 GA decays into filaments after
a propagation distance of $z=70\,\mu$m; this is seen in
figure~\ref{fig:2}(a) in terms of the B-field and in
figure~\ref{fig:2}(b) in terms of current density.
Actually we find that the filaments start to grow in the
halo region ($z<50\,\mu$m) and are then strongly amplified
in the density slopes of the blob. The growth rate is
consistent with the linear theory developed in
\cite{gremillet}. 
Resistive filamentation scales with plasma resistivity and
therefore depends strongly on electron temperature $T_e$.
In the lower density region, ohmic heating and Coulomb
energy deposition lead to high electron temperatures with
a mean value of $\langle T_e \rangle \approx 40\,$keV,
much higher than the ion temperature $T_i$, and here
magnetic fields saturate at levels of 100 T due to low resistivity.
At higher densities ($z > 70\,\mu$m), sufficient energy transfer
from electrons to ions takes place, and we find $T_e\approx T_i\leq 20\,$keV.
The corresponding ion temperature is plotted in figure~\ref{fig:3}(a).
Here higher resistivity leads to $B$-field of $\approx 1\,$kT and
enhanced filament growth. One should notice that the resistive
filamentation observed here is weaker than that obtained
from collisionless PIC simulations \cite{Honda2000,Sentoku2003}
and has a wider spatial scale. Given the broad energy
spectrum of beam electrons only those with lower energies
($< 1.5\,$MeV) are trapped in the current channels, while
the others scatter freely and tend to smooth the structure.
We find that both temperatures and $B$-fields are well described by the
analytic scaling laws derived by Bell and Kingham \cite{bell2}.
We have also checked that filamentation persists
for initial temperatures of 1 keV and halo densities up to 10 g/cm$^3$,
but disappears for beam kinetic energies $\langle E\rangle > 4.5$ MeV.

The filaments shown in figure~\ref{fig:2}(b) carry about 10~MA beam current
each. This beam current is almost completely compensated by the plasma
return current, implying $j_r\approx -j_b$; the net current
of a filament is only about 10~kA, consistent with the magnetic field
strengths of 1~kT observed in figure~\ref{fig:2}(a).
It is worthwhile emphasizing that the filamented current
distribution heats electrons significantly more than
a uniform current distribution of same total current
because of the $j_r^2$ dependence of ohmic heating.
This is how filamentation leads to enhanced beam stopping
within the present model. It differs from the strong anomalous
stopping found in PIC simulations \cite{Honda2000,Sentoku2003}
for lower plasma densities (about 10$\times$beam density).
Sentoku et al. have interpreted this stopping
as stochastic scattering of return current electrons by magnetic
perturbations $\vert B\vert$ giving rise to an additional
effective resistivity $\eta_{eff}\approx \vert B\vert/(en_p)$.
Even taking the scaling $\vert B\vert \propto n_p^{3/5}$ from \cite{bell2},
$\eta_{eff}$ decreases with plasma density $n_p$.
For the parameters of the present simulation, we find
$\eta_{eff}$ to be always smaller than the Spitzer resistivity.
Therefore, we conclude that anomalous stopping of this kind
plays no significant role here, in agreement with the results
of Mason \cite{mason}.

\begin{figure}
\begin{center}
\includegraphics[width=.45\textwidth]{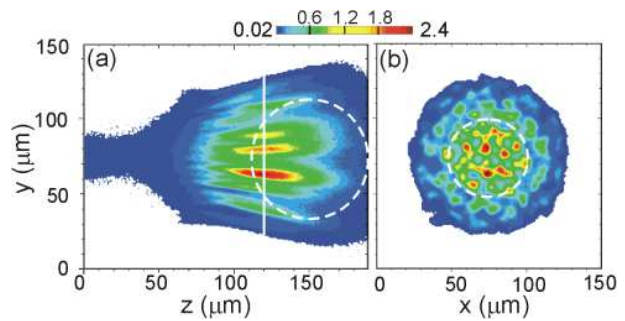}
\caption{\label{fig:4} Pressure of DT in Tbar
for the electron pulse with a mean kinetic
energy of 2.5 MeV: (a) longitudinal cut
at $x=75$ $\mu$m and (b) transverse cut at $z=120$ $\mu$m.
Plasma densities higher than 200 g/cm$^3$ are located inside
the dashed circle.}
\end{center}
\end{figure}

Figures \ref{fig:3} and \ref{fig:4} present the
central results of this paper, showing DT fuel
close to ignition. Notice that beam filamentation
in the corona is responsible for the fragmented heating
pattern in the core. Artificial suppression of the
beam-generated fields (i.e. Coulomb deposition only)
would lead to smooth core heating with a maximum
temperature of 4 keV. This is shown
in Figs. \ref{fig:3}(c) and (d), for comparison.
With fields present, a multi-hot-spot ignition
region is formed in the high-density fuel
with maximum temperatures beyond 10 keV.
The conjecture here is that this will help ignition
due to the nonlinear scaling of fusion reactivities
with temperature. Actually, this needs to be confirmed
in more detailed simulations, including hydrodynamics
and fusion reaction physics in 3D geometry.

Here we give some estimates, based on the pressure distribution
shown in figure~\ref{fig:4}. The DT ignition condition
(see page 85 in \cite{MtV2})
can be written in compact form as
$p_hR_h > 45 (\rho_h/\rho_c)^{1/2}$ Tbar$\mu$m, where
index $h$ refers to hot fuel and index $c$ to surrounding cold fuel.
This condition combines the threshold values for $\rho_h R_h$ and $T_h$
and holds for temperatures $5<T_h/$keV$<15$.
From Figs.~\ref{fig:3}(a) and (b), we find for the central hot spot
$2p_hR_h\approx 50$ Tbar$\,\mu$m in longitudinal direction. In transverse
direction, it is $2p_hR_h\approx 15$ Tbar$\,\mu$m, but here a number
of neighboring hot spots will cooperate and $2p_cR_c\approx$ 50 Tbar$\,\mu$m,
obtained from $p_c=1$ Tbar and $2R_c=$ 50$\,\mu$m, may serve as an estimate.
We conclude that the reference case shown in Figs.~\ref{fig:3} and
\ref{fig:4} is close to ignition.
It should be understood that the core heating is almost exclusively
due to Coulomb deposition of beam electrons. Ohmic heating
by return currents dominates in the halo, but plays only a minor role for
the overall energy balance. Beam-generated fields turn out
to contribute to the core heating indirectly, mediated by
filamentation and collimation effects.
Beam collimation is observed in figure~\ref{fig:3}(a) when
compared  with figure~\ref{fig:3}(c).

\begin{figure}
\begin{center}
\includegraphics[width=.35\textwidth]{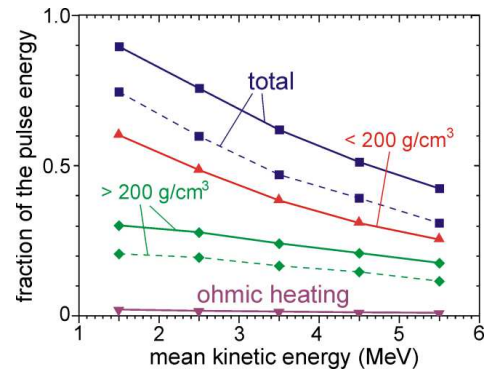}
\caption{\label{fig:5}
Fraction of total pulse energy deposited in
the target (squares), in low-density ($<$200 g/cm$^3$)
zones (triangles) and in high-density ($>$200 g/cm$^3$)
core (diamonds). Solid lines correspond to full simulations
with beam-generated fields and Coulomb energy
deposition. Dashed lines correspond to simulations
with beam-generated fields suppressed. The fraction
due to ohmic heating is small throughout.}
\end{center}
\end{figure}

Figure~\ref{fig:5} shows what fractions of the injected beam energy
are deposited in different parts of the target. The total deposition,
given versus beam kinetic energy $\langle E \rangle$, drops from 90\% to 40\%,
when raising $\langle E \rangle$ from 1.5 to 5.5 MeV. Most of the
deposition is due to classical Coulomb collisions, consistent with
the areal density of 2.9~g/cm$^2$ along the axis.
Notice that the other part of the energy is not deposited
at all, but passes through the target and is lost. Clearly this
makes average beam energies beyond 5 MeV prohibitive. The very
important partition between deposition into high density core and
low-density zones with $\rho <$ 200 g/cm$^3$ is also shown in
figure~\ref{fig:5}. We find that the energy coupling to the core
amounts to 30\% at 1.5 MeV and 20\% at 5 MeV. It is less sensitive
to $\langle E \rangle$ than the coupling into the lower density
regions. Of course, the core coupling strongly depends on the
divergence angle of the injected beam, which is therefore a key
parameter for fast ignition. In the present simulations, the
core coupling efficiency degrades by 40\%, when raising the angle
from 22.5$^\circ$ to 30$^\circ$. On the other hand,
one should notice that magnetic pinching of the beam
improves core coupling significantly. Suppressing all
beam-generated fields in the simulations would lead to
the dashed curves in figure~\ref{fig:5}.

Estimating the laser-to-fast-electron conversion efficiency to be 50\%,
we find a laser pulse energy of 100 - 150 kJ necessary to ignite a target.
There may be possibilities to reduce this energy,
e.g. by shortening the distance between cone tip and blob or by careful
design of the cone to reduce beam divergence \cite{Kodama2004}.
Certainly, transport in the cone needs to be included in more
complete studies, in particular to account for the potential barrier
\cite{mason} and filamentation \cite{Sentoku2004} at the cone tip.

In conclusion, the message of this paper concerning fast
ignition of inertial fusion targets is that a giga-ampere, multi-PW
current can be transported through the steep gradients of the
plasma corona toward the high-density fuel core. This is shown
here for the first time in 3D geometry, using hybrid PIC
simulation. Central questions could be answered: Collective
magnetic effects play a major role for core heating, but in an
indirect way. Resistive beam filamentation grows in the
low-density halo and seeds the 3D multi-prong beam, which then
penetrates the core. Of course, 3D simulation is crucial in this
context.

In the core, collective behavior is suppressed due the large
plasma-to-beam density ratio, and energy deposition takes place
almost exclusively by classical Coulomb collisions.
We find a fragmented hot spot configuration, and
the fragmentation may actually help fuel ignition, since
concentrating the energy in a number of prongs
rather than heating the whole volume spanned by the prongs leads
to higher temperatures and therefore to more heating by fusion
products. The $\langle$pR$\rangle$ values obtained for
the reference case indeed indicate that this case is close to ignition.
More detailed simulations including hydrodynamics and fusion
heating are now in progress to confirm this point.

Concerning collective beam deposition, we find, within
the physical model used here, that indeed beam filamentation
enhances ohmic heating, because it depends quadratically
on the return current density $j_r$. But
this additional deposition is not identical
with the anomalous stopping found in PIC simulations at lower densities.
It also contributes little to the overall energy balance
in the present simulation of fast ignition.
Rather the self-generated B-fields help by
collimating the relativistic beam, and this improves the coupling
efficiency substantially.

\begin{acknowledgments}
This work was supported by the research grant FTN2003-6901
of the Spanish Ministry of Education and by the Association
EURATOM - IPP Garching in the framework of IFE Keep-in-Touch
Activities and the Fusion Mobility Programme.
\end{acknowledgments}

\end{document}